\documentstyle[psfig]{mn}
\ifnfsstwo

\fi 

\ifnfssone 
  \newmathalphabet{\mathit} 
    \addtoversion{normal}{\mathit}{cmr}{m}{it} 
    \addtoversion{bold}{\mathit}{cmr}{bx}{it}

\fi 
 
\ifoldfss

\fi 
 
\loadboldmathitalic 
\loadboldgreek

\begin{document}

\title{New diagnostic methods for emission-line galaxies in deep 
surveys} 
\author[Rola, Terlevich \& Terlevich] 
       {Cl\'audia S. Rola$^{1,2}$, Elena Terlevich$^{1}$ \& Roberto J. Terlevich$^{2}$ 
        \\  
	$^{1}$ Institute of Astronomy, University of Cambridge, 
Madingley Road, Cambridge CB3 0HA, United Kingdom \\ (crola {\it or} et @ast.cam.ac.uk)\\
	$^{2}$ Royal Greenwich Observatory, Madingley Road, 
Cambridge CB3 0EZ, United Kingdom  \\ (rjt@ast.cam.ac.uk) \\
        } 
	\date{Accepted March 1997, Received November 1996} 
	\pubyear{1997}

\maketitle 
 
\begin{abstract} 

We present  new  quantitative classification methods for emission-line
galaxies, which are specially
designed to be used in deep galaxy redshift
surveys.  A good segregation between starbursts and active galactic nuclei,
i.e.~Seyferts 2 and LINERs, is obtained
from diagnostic diagrams involving the  [O II]$\lambda $3727 $\AA$, [Ne III]$\lambda
$3869 $\AA$, H$\beta$  and [O III]$\lambda$5007 $\AA$\
relative intensities or the [O II]$\lambda $3727 $\AA$\ and H$\beta$ equivalent 
widths.
Furthermore, the colour index of the continuum underlying [O II]$\lambda $3727 $\AA$\ and H$\beta$ 
provides an additional separation parameter between 
the two types of emission-line galaxies.

We have applied the equivalent widths method to the $0 < z \leq 0.3$
emission-line galaxies of the Canada-France Redshift Survey. 
Our results are in very good agreement
with those obtained using the standard diagnostic diagrams including all
the strong optical emission-line intensity ratios.

\end{abstract} 
 
 
\section{Introduction} 

Deep redshift surveys have produced in the past 
few years a relatively large number of optical/near UV spectra 
of galaxies at redshifts $z < 0.7-0.8$ (e.g., Dressler \& Gunn 1983;
Couch \& Sharples 1987; Broadhurst, Ellis \& Shanks 1988; Lavery 
\& Henry 1988; Colless et al. 1990; Colless {\it et al.} 1993; 
Songaila {\it et al.} 1994; Glazebrook {\it et al.} 1995; 
Le F\`evre {\it et al.} 1995).
A large percentage of galaxies in these deep surveys present 
narrow emission lines.
The analysis of the spectrum of these distant emission-line galaxies 
(hereafter ELGs)  should
provide important information about their intrinsic properties 
(stellar population, rate of star formation, metallicity, etc.) 
and the evolution of these parameters with increasing look-back time.
But a central problem is that very little is known about the nature, either 
H II galaxies
or active galaxies\footnote{In this paper, we will call active galaxies those
hosting an active galactic nucleus (AGN). These are Seyferts 2 and LINERs 
-- Low Ionisation Narrow Emission-line Regions -- and we excluded  Seyferts 1 
as these are generally easier to identify
because they present broader Balmer lines (FWHM $>$ 1, 000 Km s$^{-1}$).}, of 
the ELGs  discovered in these deep surveys. 

In H II galaxies the emission-line spectrum is dominated
by the emission originating in H II regions, where 
ultraviolet photons emitted by OB stars ionise the surrounding gas, while in 
the case of active galaxies
the gas ionising source is much harder and has the shape of a power law.
Furthermore, star forming regions are always associated 
with H II regions and notably with H II galaxies, while that is not necessarily
the case for active galaxies. By determining the
different nature (active or H II galaxy) of the ELGs observed in redshift 
surveys, one can obtain a more precise physical picture
of the galaxy evolution process.

Nearby ELGs are classified  
using the emission-line ratios of the most prominent optical lines like: 
[O I]$\lambda$6300 $\AA$, [O II]$\lambda\lambda$3727, 3729 
$\AA$, [O III]$\lambda\lambda$4959, 5007 $\AA$, [N II]$\lambda$6584 $\AA$,
[S II]$\lambda\lambda$6717, 6730 $\AA$, H$\alpha$ and H$\beta$  
(see Baldwin, Philips \& Terlevich 1985; Veilleux \& Osterbrock 1987). 
In higher redshift objects, however, most of these lines move out of the 
readily observable optical spectral range 
making impossible the use of current methods to classify 
ELGs  beyond $z \approx 0.3$.
For $z > 0.3$ the only strong emission-lines remaining in 
the optical range are H$\beta$ and [O III]$\lambda\lambda$4959, 5007 $\AA$,
observable up to $z \approx $ 0.7, 
and [O II]$\lambda\lambda$3727, 3729 $\AA$\ and  [Ne III]$\lambda$3869 $\AA$\
observable up to 
$z \approx $ 1.2.
Typical deep redshift survey spectra have a low signal-to-noise ratio,
so they usually show only
the two strongest emission lines, like [O II]$\lambda$3727 $\AA$\ \footnote{Hereafter, 
[O II]$\lambda$3727 $\AA$\ represents the sum of the lines 
[O II]$\lambda\lambda$3727, 3729 $\AA$.} 
and H$\beta$, rendering the existent methods of classification useless. 
Nevertheless, in spite of not being able to determine their nature, the large 
values of EW([O II]) observed in ELGs at  
intermediate redshifts lead researchers to suggest 
high rates of present star formation and therefore
the existence of a significant fraction of starburst galaxies (e.g., 
Broadhurst {\it et al.} 1988, Colless {\it et al.} 1990).

However, it is important to remember that the intensity or the 
equivalent width of an emission-line is related to the present star 
formation rate {\it only} if the ionisation {\it and} observed
continuum sources are OB stars as is the case in
H II galaxies (i. e., starburst galaxies, blue compact galaxies, etc), 
which is not necessarily the case for galaxies hosting an  
AGN (i.e.~ Seyferts type 1 and 2, and LINERs).

\begin{figure*}
\centerline{\psfig{file=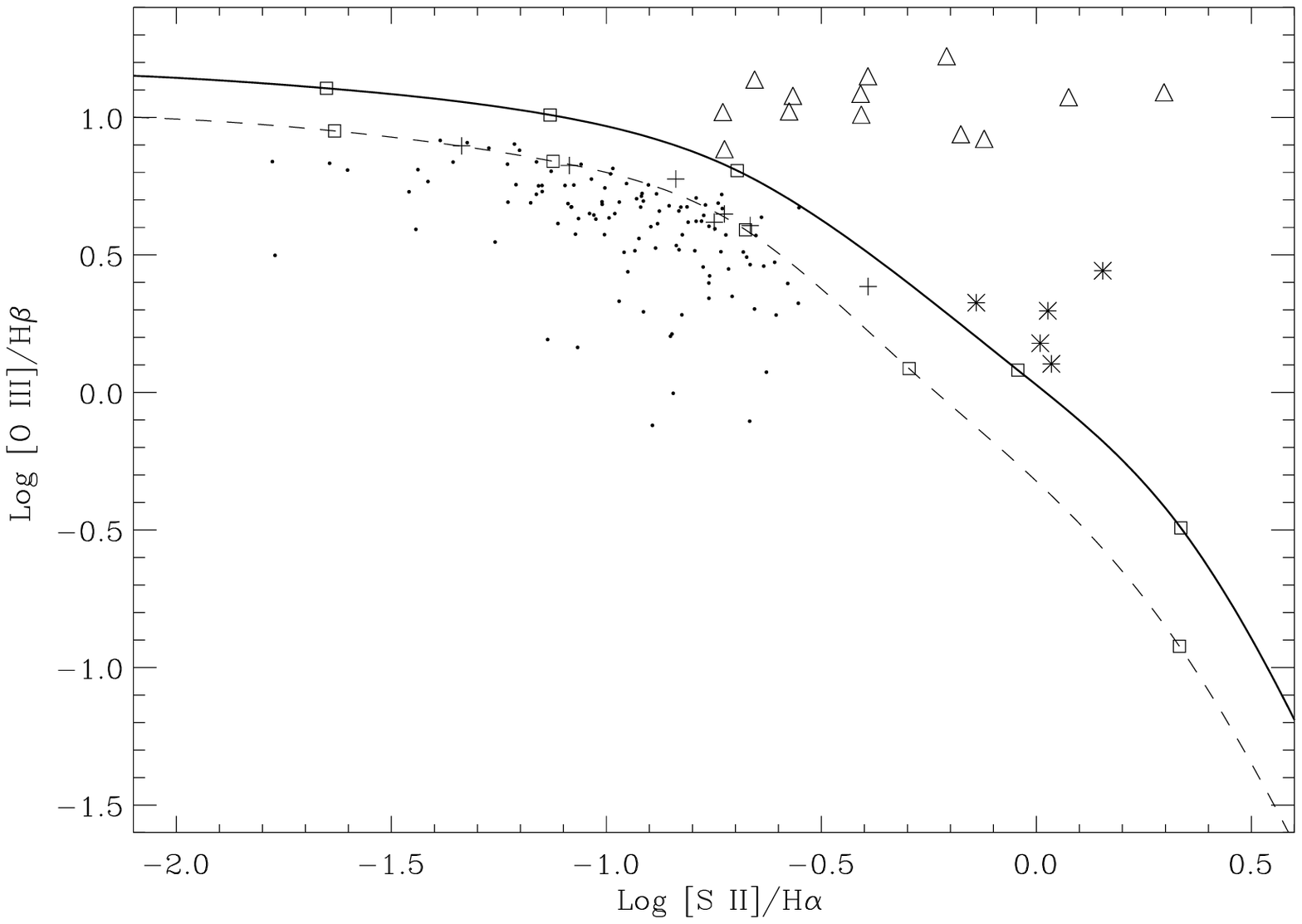,width=9.0cm,height=9.0cm}
\hspace{-0.4cm}\psfig{file=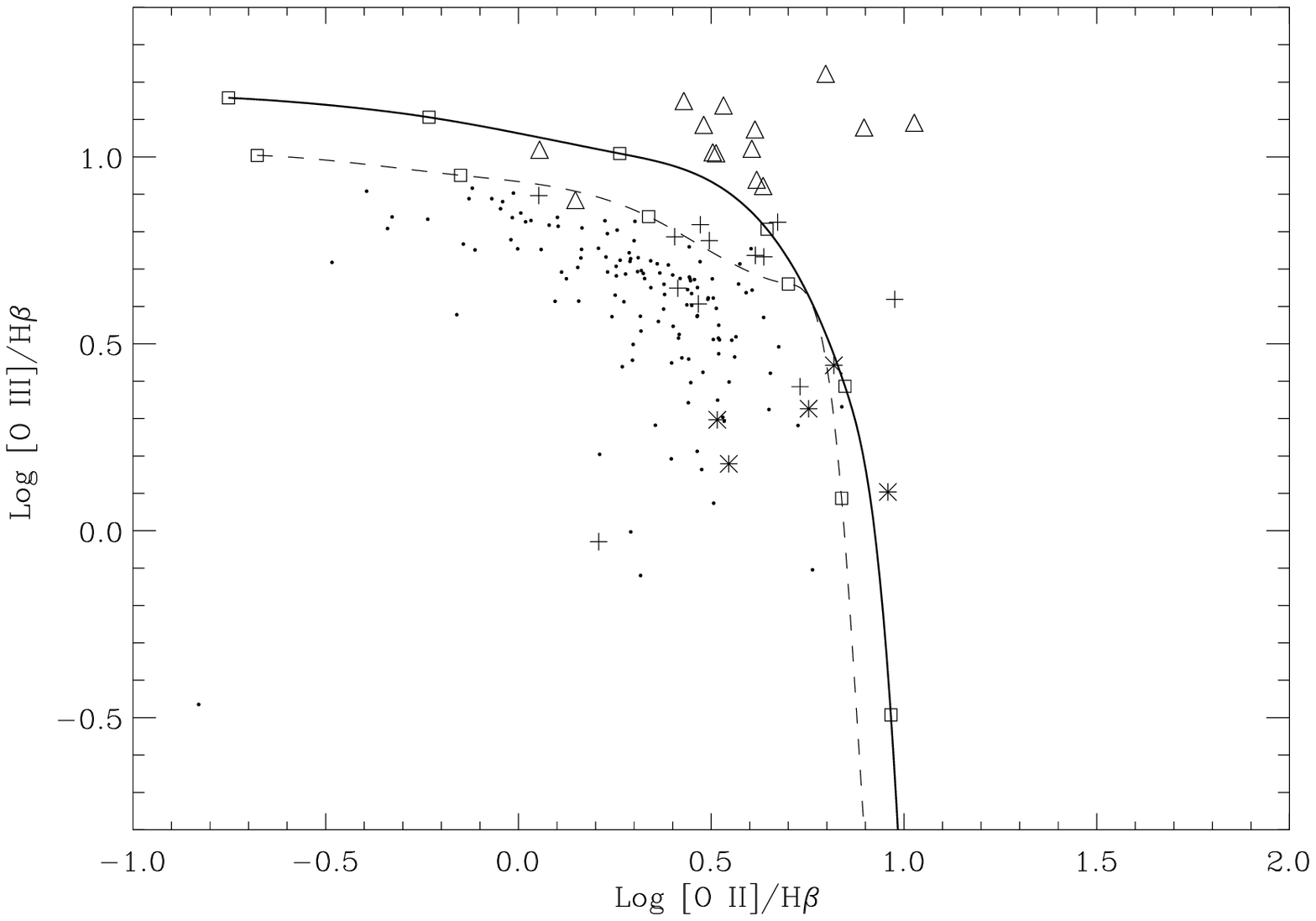,width=9.0cm,height=9.0cm}}
\vspace{-0.4cm}
\centerline{\psfig{file=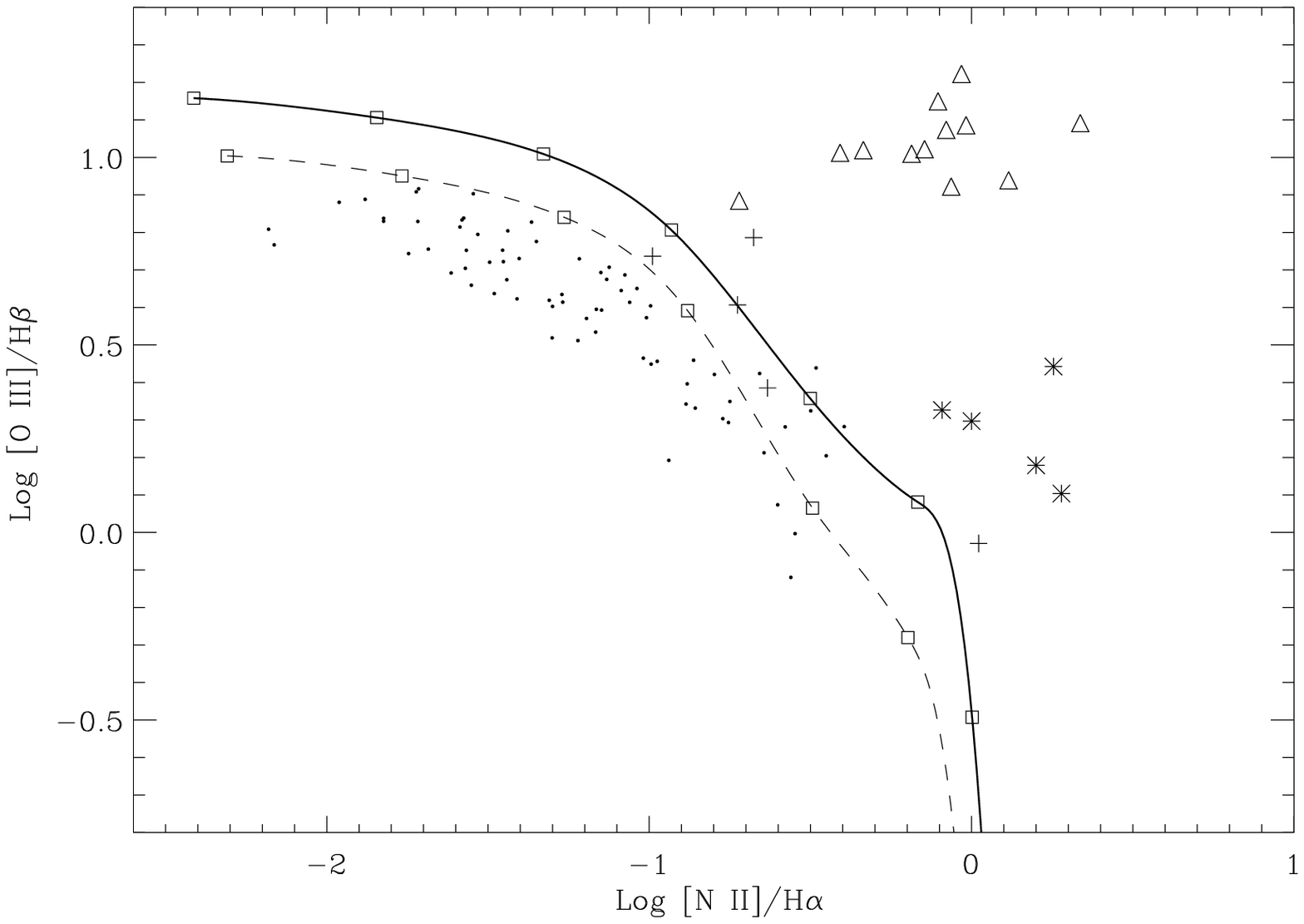,width=9.0cm,height=9.0cm}
\hspace{-0.4cm}\psfig{file=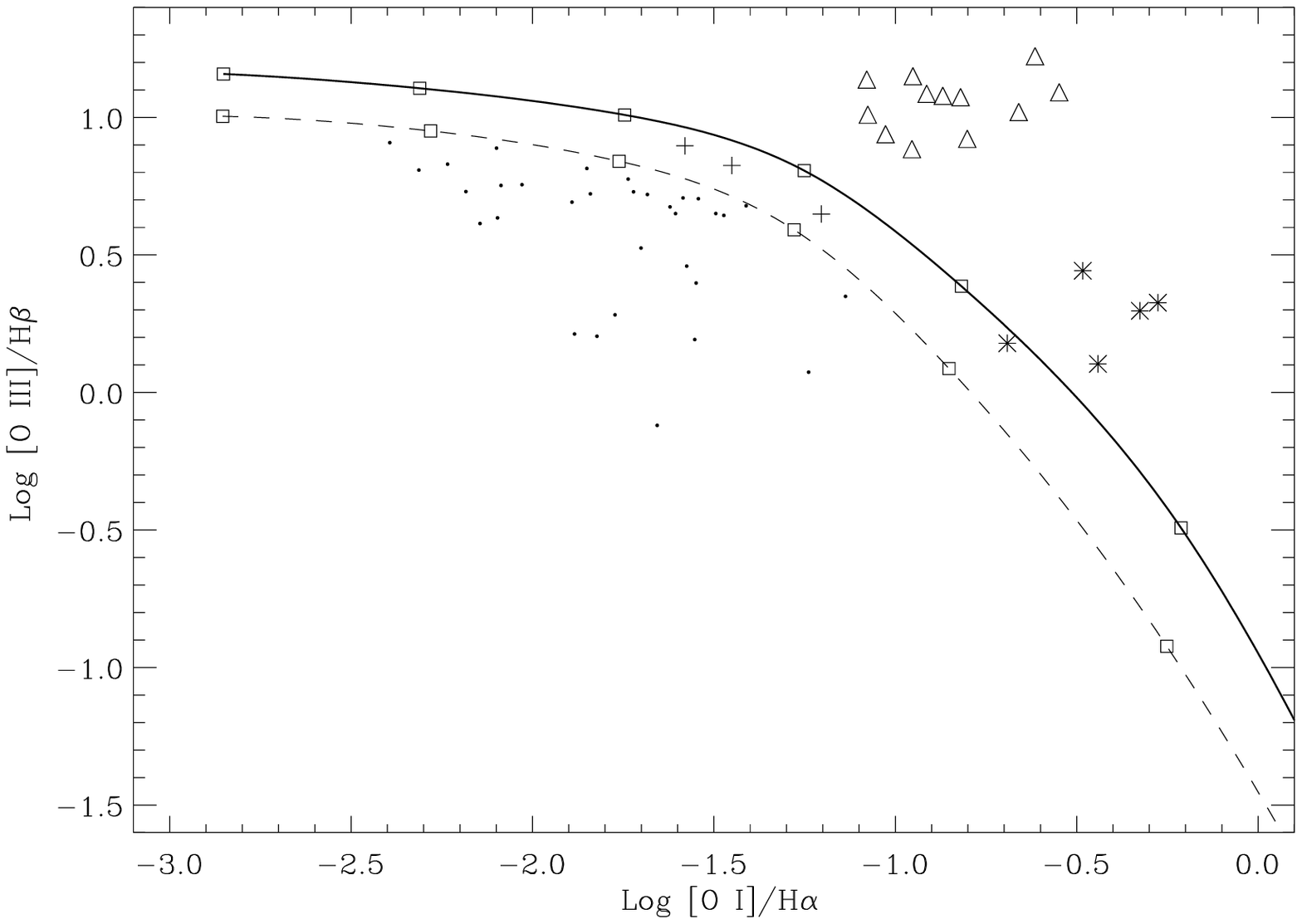,width=9.0cm,height=9.0cm}}
\vspace{-0.3cm}
\caption{\label{f0} The four diagnostic diagrams used in the classification 
of the reference sample. The curves 
correspond to photoionisation by OB stars, for stellar effective 
temperatures of 50,000 K (dashed line) and  60,000 K (full line) respectively
(see text). The ELGs classified as Seyfert 2 are represented by triangles, 
LINERs by asterisks and the dots are H II galaxies; crosses mark the 
transition or ambiguously classified objects.}
\end{figure*}

Tresse {\it et al.} (1996) analysed the Canada-France 
Redshift Survey (CFRS) galaxies up to a redshift of 0.3 and found
that about 17 $\%$ of all objects have emission-line ratios consistent with being active galaxies (like Seyferts 2 or 
LINERs). Unlike in previous deep redshift surveys, a 
classification was possible thanks to the wide 
spectral range observed (4500 - 8500 $\AA$) 
which allowed to use the classical diagnostic diagrams (see also Rola, 1995).
Their result indicates that in deep surveys the 
fraction of ELGs that are active galaxies may be higher 
than previously thought. It becomes therefore very
important to  compare the CFRS results with other surveys and 
to extend this analysis to higher redshifts. 
To accomplish this it is necessary first to  
find new diagnostic methods that are appropriate for
deep surveys and could properly separate  active galaxies from 
H II galaxies at $z >$ 0.3.

In this paper we 
investigate new methods of quantitative classification that involve only 
the strong emission-lines that can be detected in deep redshift surveys,  
as well as the continuum colour. As reference we used a large sample 
of nearby ELGs, comprising 231  high quality spectra of H II galaxies, Seyfert 
2 nuclei and LINERs, classified combining the usual diagnostic diagrams 
method with an extended grid of photoionisation models. Section~2
explains the methodology used.  Section~3
describes the characteristics of the sample of ELGs
used in our study and explains how the 
classification of the reference local sample of ELGs was made.
Section~4 presents the new diagnostic 
diagrams and discusses the advantages and the problems of the new methods.
Conclusions are drawn in section~5.

\section{Methodology}

The main steps in the methodology used in this work are the following.
We have first compiled a quality sample of local emission-line spectra
in which we performed a reliable classification of the nature
of the ELGs. This classification is 
based on a purpose built grid of photoionization models,
allowing an improved approach compared with the usual empirical
methods.

After classifying the nearby sample into 3 groups
of objects: definite H II galaxies, definite active galaxies 
(Seyferts 2 and LINERs) and intermediate objects,
we analysed the definite H II galaxies and active galaxies sets,
in order to search for new diagnostics to identify the nature of 
ELG's involving {\bf only} lines to the blue of  5000 $\AA$, approximately. 
This is because we want the new diagnostic methods to be applicable
to optical redshift surveys up to $z$ $\approx$ 0.75.
In order for the new methods to be also applicable to spectra
which are not flux calibrated we have built some diagnostic methods
based only on the equivalent widths of certain emission-lines.
For the flux calibrated data, the new methods use also emission-line
intensity ratios and/or the continuum intensity.

While building this compilation of emission-line spectra
we have examined all the published data on H II galaxies
and active galaxies. 
As we are interested in the nature segregation, 
we used only the best available optical data in terms of emission-line 
spectra. Therefore, we opted to use spectra with good data in the
region 3600 $\AA$ to 5100 $\AA$. Nevertheless, this is not often 
easy to obtain.
To begin with, until very recently only
a few detectors/spectrographs had high sensitivity
in the region from 3600 $\AA$ to 4000 $\AA$.
On top of that, most observers do not publish
the equivalent widths of the strongest lines, so 
the information about the continuum shape is lost.

\section{Reference sample: characteristics and nature classification}
\label{sample}

The data sample we used in this work comes mainly from the 
Terlevich {\it et al.} (1991) spectrophotometric catalogue of H II galaxies 
(where some  AGNs are also included), selected using slitless spectroscopy. 
Additional ELGs come from the Cambridge survey (see Masegosa {\it et al.} 
1994). The data for typical AGN's was augmented using published
spectrophotometry from Ho {\it et al.} (1993) and 
Costero \& Osterbrock (1977). 
The condition (trivial) imposed for an object to be included in the
final sample was that the data had to contain the lines to be used in the new 
methods, i. e., at least [O II]$\lambda$3727 $\AA$, 
[Ne III]$\lambda$3869 $\AA$, H$\beta$ and [O III]$\lambda$5007 $\AA$\ and the equivalent 
widths of [O II]$\lambda$3727 $\AA$\ and H$\beta$. One large problem found is that 
there 
is in the literature only a handful of AGNs with measurements of 
the equivalent widths of [O II]$\lambda$3727 $\AA$\ and H$\beta$.

\begin{figure}
{\psfig{file=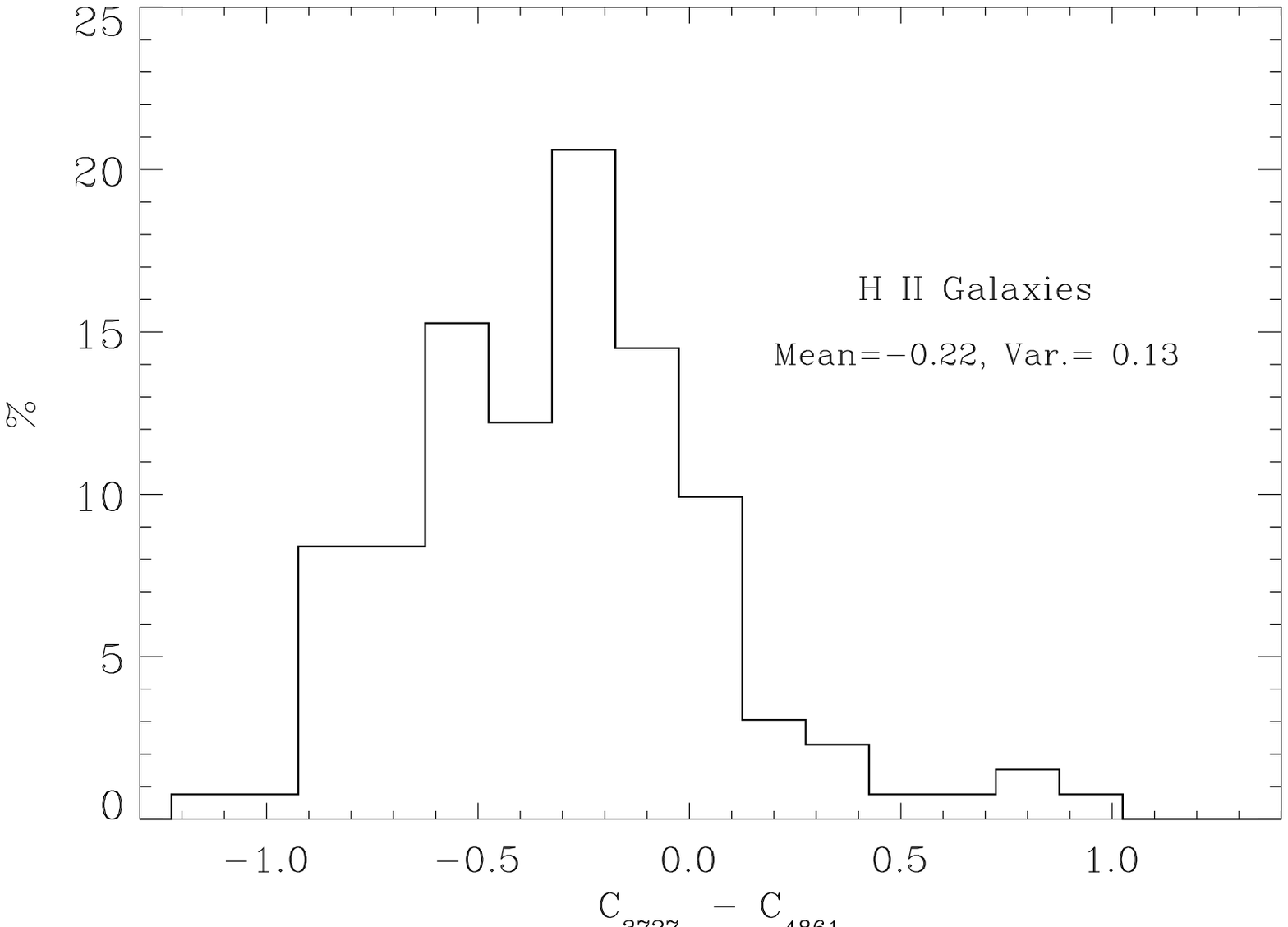,width=9.0cm,height=7.0cm}}
\vspace{-0.2cm}
{\psfig{file=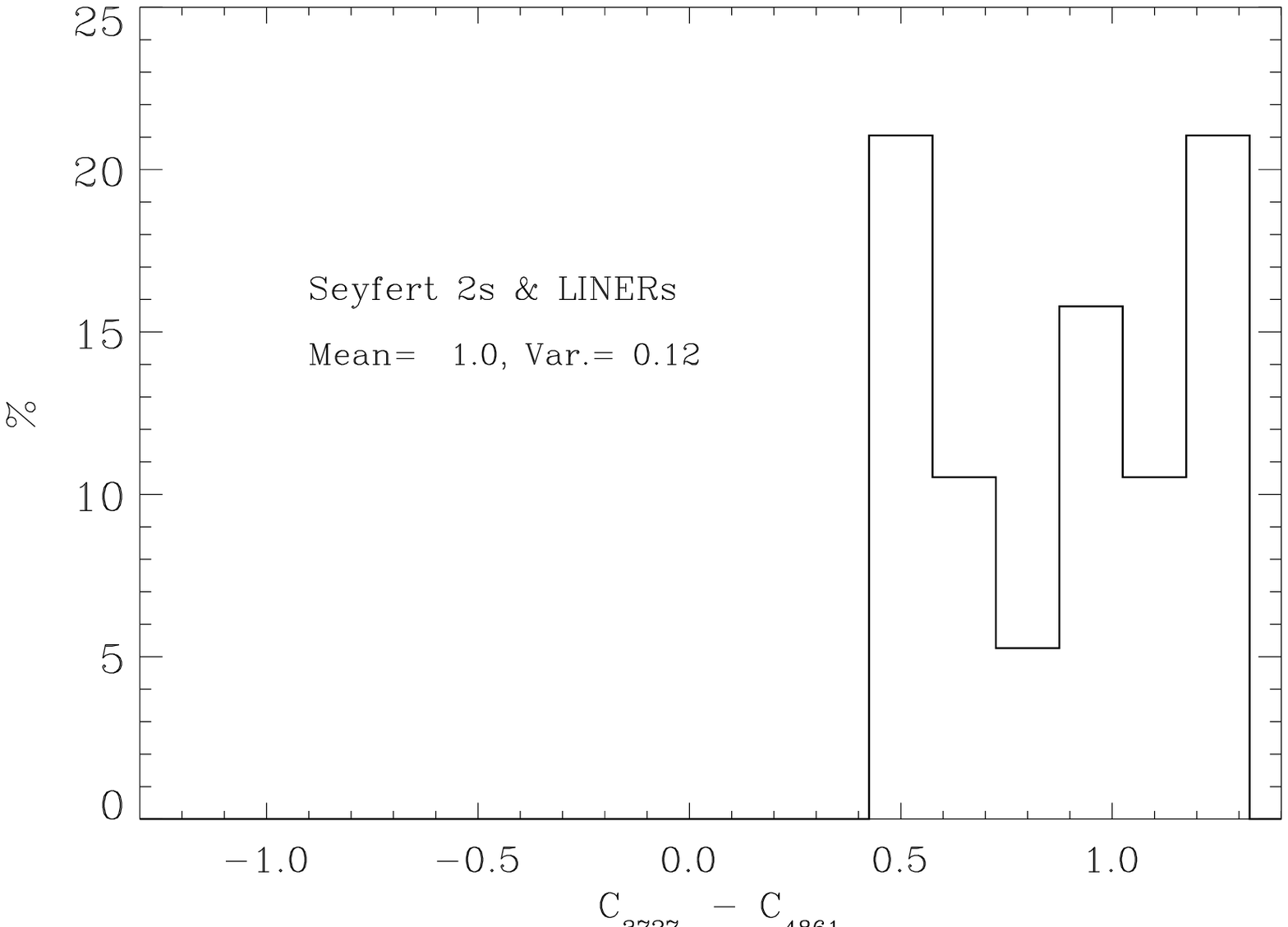,width=9.0cm,height=7.0cm}}
\vspace{-0.2cm}
\caption{\label{f1} The distribution of the $C_{3727} - C_{4861}$ colour 
index is shown for H II Galaxies (top) and for active galaxies (bottom). The 
percentages are relative to the total number of ELGs of each type. 
The mean and variance of each distribution are shown in the diagrams.}
\end{figure}

For the classification of the sample four diagnostic diagrams were 
used:  [O III]$\lambda$5007/H$\beta$ versus 
[S II]$\lambda$6725/H$\alpha$\footnote{[S II]$\lambda$6725 $\AA$\ 
represents the sum of the lines [S II]$\lambda\lambda$6717, 6731 $\AA$}, 
[O III]$\lambda$5007/H$\beta$ versus 
[N II]$\lambda$6583/H$\alpha$, [O III]$\lambda$5007/H$\beta$ 
versus [O I]$\lambda$6300/H$\alpha$   and [O III]$\lambda$5007/H$\beta$ 
versus [O II]$\lambda$3727/H$\beta$\footnote{intensity ratio corrected for 
reddening} (see Baldwin, Philips \& Terlevich 1985, Veilleux 
\& Osterbrock 1987 and Rola 1995).  These are presented in figure~\ref{f0}
with the curves determined from the photoionisation models,  separating
the H II galaxies region from the active galaxies region (see following 
section for details).

\subsection{Photoionisation models}

We have determined the boundary between H II galaxies
and active galaxies in the diagnostic diagrams using an
extensive grid of steady-state spherically symmetric 
 H II region photoionization models computed with
the code PHOTO (see Rola 1995 and Stasi\'nska 1990).  
To mimic the complexity of  H II galaxies and best determine their loci 
in the diagnostic diagrams, we have considered wide ranges for the 
physical parameters of the models.

There are five main parameters driving the photoionization models used to 
determine the loci of H II galaxies in the diagnostic diagrams: the hydrogen 
density ($n_{H}$), the metallicity ($Z$), the number of ionizing stars 
($N$), the filling factor ($f$) and the ionizing source effective 
temperature (T$_{\rm eff}$). All the grid models considered a 
constant hydrogen density, $n_{H}$=10 cm$^{-3}$. Varying $n_{H}$ 
in the range found in H II regions ($n_{H}$=10 $-$ few 100 cm$^{-3}$) has a 
negligible effect on the intensity ratios. Models are dust free (the effects 
of dust were considered only as depletion) and were calculated using
three free parameters, 
the ionisation parameter, $ U = Q_{H} /(4 \pi R^{2} n_{H} c)$ (where 
$Q_{H}$ is the number of $H^{o}$ ionising photons,  
R is the radius of the photoionised region and c is the speed of 
light), the effective temperature of the ionizing radiation, T$_{\rm eff}$, 
and the metallicity of the gas, $Z$.
$U$, which can also be written as
$ U \propto (n_{H} f^{2} N Q_{H}^{\star})$ (where $Q_{H}^{\star}$ is 
$Q_{H}$ for a single star and $N$ is the number of ionising stars) varied 
roughly between
0.2 and $10^{-6}$. For simplicity, we have considered the ionising
source as a ``cluster'' with a single type of stellar atmosphere, 
with T$_{\rm eff}$
ranging from 3 $\times$ $10^{4}$ K to 6 $\times$ $10^{4}$ K. This is 
a reasonable approximation if the hottest stars are assumed to dominate 
the total flux of ionising photons. For the 
distribution of the ionising radiation we have used  the $\log g= 5.0$ 
Kurucz model atmospheres (Kurucz, 1992) with an 
abundance consistent with the one in the nebula. We considered 
four metallicities: 2 $Z_{\odot}$, $Z_{\odot}$, 0.25 $Z_{\odot}$ and 
0.1 $Z_{\odot}$.  
More details about the grid are given in Rola (1995) and in Tresse {\it et al.}
(1996).

\begin{figure*}
\centerline{\psfig{file=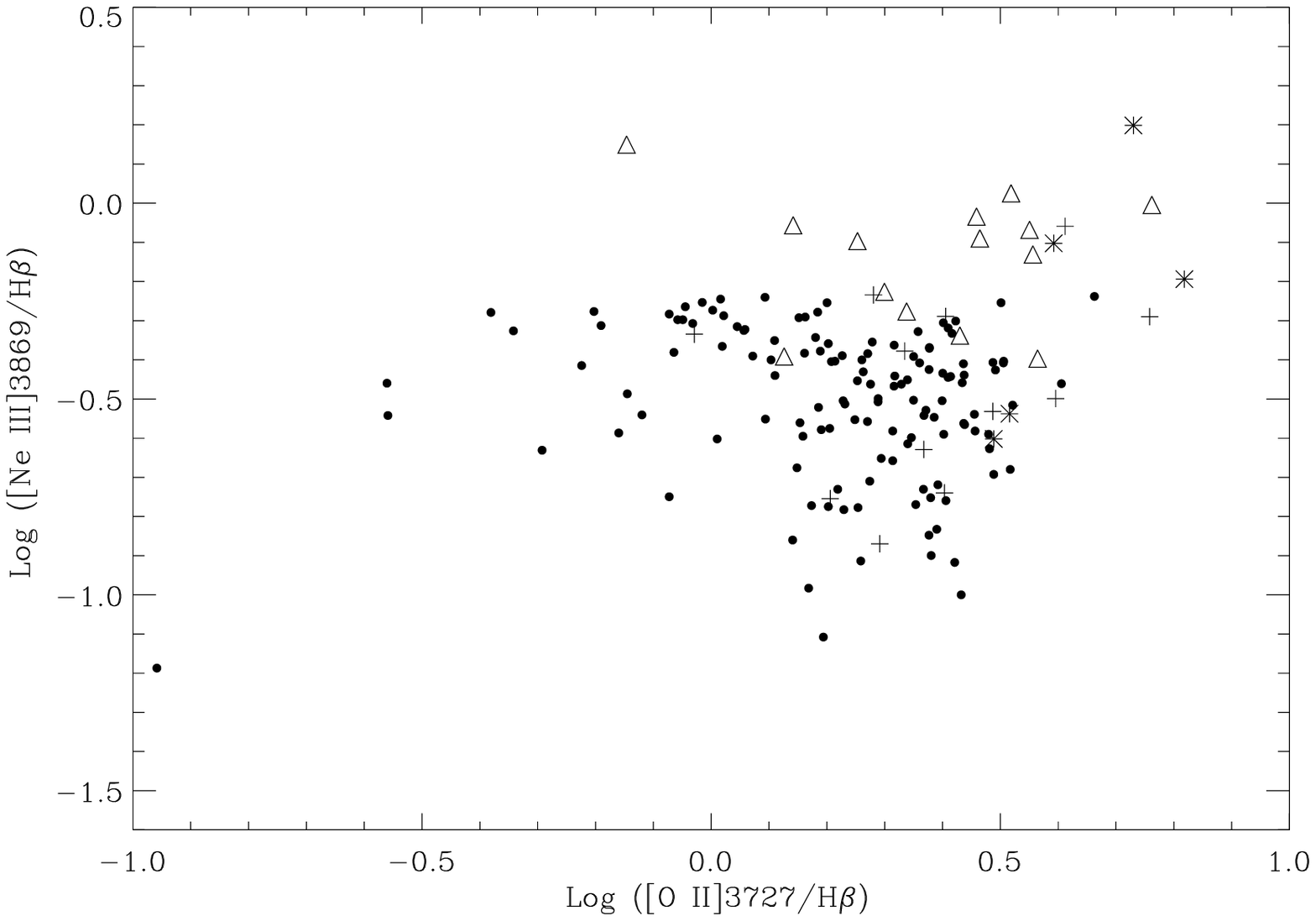,width=9.0cm,height=9.0cm}
\hspace{-0.3cm}\psfig{file=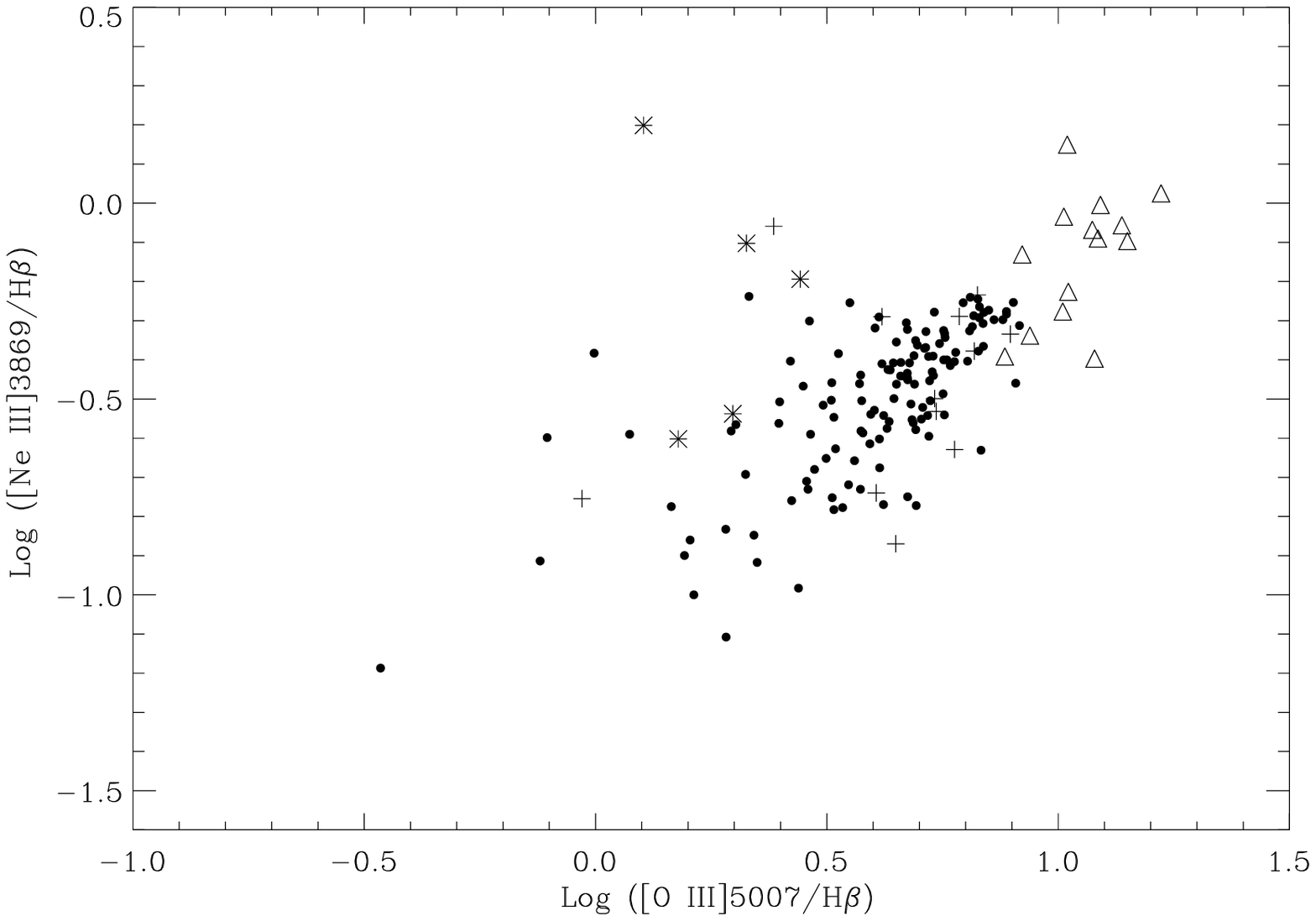,width=9.0cm,height=9.0cm}}
\vspace{-0.3cm}
\caption{\label{fa} Symbols as in Figure~1. 
{\it Left diagram}: [O II]$\lambda$3727/H$\beta$ versus  
[Ne III]$\lambda$3869/H$\beta$ intensity ratios. Note that contrary to 
the top-right diagram of figure~1, the data are not corrected for reddening.
 There is an upper limit beyond which
no more H II galaxies from our sample are observed. This is situated at about
 $ \log$ ([Ne III]$\lambda$3869/H$\beta$) $\approx$ - 0.2. 
{\it Right diagram}: [O III]$\lambda$5007/H$\beta$ versus  
 [Ne III]$\lambda$3869/H$\beta$ intensity ratios (not reddening corrected). 
The separation limit between H II and active galaxies is the same, but here the
[O III]$\lambda$5007/H$\beta$ intensity ratio allows to identify LINERs 
($\log$~([O III]$\lambda$5007/H$\beta$) $\leq $ 0.5) from Seyferts 2 
($\log$ ([O III]$\lambda$5007/H$\beta$) $> $ 0.5).}
\end{figure*}

\subsection{Boundaries between H II galaxies and active galaxies}
\label{boundaries}

Uncertainties in the input parameters of the  photoionization models
such as the stellar atmospheres spectrum and the value of the upper limit 
for the stellar 
effective temperature of the hottest stars, as well as in the values 
of atomic coefficients  used in the photoionisation models, mean that
the boundary separating H II galaxies from active galaxies cannot be
accurately defined.
Furthermore, H II galaxies are more complex systems than a
H II region with a single type of ionising stellar atmosphere, so these 
boundaries 
should be considered with some caution.
We have opted for a conservative approach and  selected the models
that define the limits to the location of H II galaxies in the 
diagnostic diagrams in a way that slightly overestimates
the region occupied by them (therefore, slightly underestimating 
the active galaxies region).  

Two boundary lines  were used to segregate the H II  galaxies
from the active galaxies by their emission-line intensity ratios. They define 
the ``extreme'' limits of H II galaxies in the diagnostic diagrams and are 
obtained  from photoionisation models with T$_{\rm eff}$ $=$ 50,000 K or 
60,000 K, $Z$ $=$ 0.25 and 1 $Z_{\odot}$,  
with $U$ varying between 2.0 $\times$ 10$^{-6}$ and 8.0 $\times$ 10$^{-2}$, 
approximately\footnote{Blackbody models with T$_{\rm eff}$ 
from 100,000 to 150,000 K 
do not change significantly the location of the H II galaxies upper envelope}. The models are shown in figure~\ref{f0}, 
where the symbols associated with each ELG represent the final classification.

\subsection{Sample classification}

We have determined the nature of the ELGs in the reference sample  
considering both T$_{\rm eff}$ $=$ 50,000 K and T$_{\rm eff}$ $=$ 60,000 K 
boundaries. ELGs in the left part of the diagnostic diagrams are H II 
galaxies while those in the right are AGNs.
Objects which fell between both separation curves 
in the diagnostic diagrams were considered {\it transition} objects. 
This final classification was obtained by
considering the balance of the 4 diagrams classification. 
If an ELG was classified both as an H II and an AGN in an equivalent number 
of diagrams, its final classification was considered  
ambiguous, and we called these also {\it transition} objects. 
Finally, among AGNs we classified as LINERs those with  
log~[O III]$\lambda$5007/H${\beta}$ $\leq$ 0.5 
(Veilleux \& Osterbrock 1987, Filippenko \& Terlevich 1993). We remark that 
our classification criteria for LINERs has been tested and seems equivalent 
to the original Heckman (1980) criteria, i.e., 
[O II]$\lambda$3727/[O III]$\lambda$5007 $\geq$ 1, and 
[O I]$\lambda$6300/[O III]$\lambda$5007 $\geq$ 1/3.

Out of the 162 ELGs that meet our selection conditions, we classified 131 as 
H II galaxies,  
14 as Seyferts 2 and 5 as LINERs. 12 objects remained unclassified due to 
ambiguity in their location in the diagnostic diagrams of Figure~1, i.e., 
either the classification was not conclusive (they appear 
as H II galaxies and as AGNs in an equivalent number of
diagnostic diagrams) or they were located between the
50, 000 K and the 60, 000 K curves  in most of the 
diagnostic diagrams of figure~1.

\section{New diagnostic diagrams}

For the emission-line reference sample discussed in section~\ref{sample}, 
a minimum of four emission-line intensities (of which two were  
H$\alpha$ and H$\beta$)  
were available for the classification of each object. This is usually not 
the case in most deep redshift surveys. 
In objects with redshift larger than 0.3, the H${\alpha}$ 
line falls out of the spectral range of CCD detectors
making reliable reddening corrections (as well as object classifications)
exceedingly difficult. An other strong emission-line commonly present in ELGs
spectra, [O III]$\lambda$5007 $\AA$, is lost  at z $>$ 0.7, approximately.
It is therefore important to be able to determine the nature of
redshifted ELGs independently of reddening 
and using a minimum number of emission-lines between [O II]$\lambda$3727 $\AA$\
and [O III]$\lambda$5007 $\AA$.
Hence we investigated diagnostic diagrams using the
[O II]$\lambda$3727 $\AA$,  [Ne III]$\lambda$3869 $\AA$, H$\beta$ 
and [O III]$\lambda$5007 $\AA$\ emission-lines and not requiring 
reddening or stellar absorption corrections. 
These are the easiest lines to detect in 
the optical range up to $z$ $\approx$ 0.75. We 
decided to use also equivalent widths (which are 
essentially reddening independent and do not require flux calibration) 
instead of using intensity ratios only. In addition  
we have investigated the use of the colour of the underlying continuum as a 
classification parameter. {\it None of the data in any of the new diagrams 
is reddening or stellar absorption corrected}, so that we would be working in
the same conditions with the local sample of ELGs as we would encounter
in a  redshifted one.

\subsection {The blue continuum colour}

We have defined the continuum colour index ``$C_{3727} - C_{4861}$'' as:

\begin{equation}
C_{3727} - C_{4861} = 2.5 \times \log  \left( \frac{C_{H\beta}}{C_{[O 
II]\lambda 3727}} \right) ,
\end{equation}

\noindent
where $C_{H\beta}$ is the intensity of the continuum underlying
H$\beta$.
Accordingly with the parameters available from our data this
was calculated as:
 
\begin{equation}
C_{3727} - C_{4861} = 2.5 \times \log  \left( \frac{EW({[O II]\lambda 
3727})}{EW({H\beta})} \times \frac{I_{H\beta}}{I_{[O II]\lambda 
3727}} \right)
\end{equation}

\noindent 
where the $ \frac{I_{H\beta}}{I_{[O II]\lambda 3727}}$  flux calibrated 
intensity ratio has not been  
corrected from reddening. With the above definition,  
larger $C_{3727} - C_{4861}$ values correspond  to redder continua.

The distribution of this colour index for our ELGs sample is shown in 
figure~\ref{f1}, where it can be seen that AGNs tend to have higher values of  
$C_{3727} - C_{4861}$, i.e.  a redder continuum,  than H II galaxies.   
This is potentially an important result for the classification of ELGs;  
all AGNs in figure~\ref{f1} have $C_{3727} - C_{4861}$ $\geq $ 0.4 against 
only about  4$\%$ of all H II galaxies.

This dichotomy in the colour index distribution is probably a reflection
of the fact that in the local Universe while most starburst galaxies have
late Hubble types (Sc-Sd),
galaxies with AGN tend to be of early Hubble type (Sa-Sc) (Terlevich {\it et 
al.} 1987). While red giants contribute significantly to the 
continuum of the central regions of galaxies with an AGN, 
young OB stars are the main contributors to the optical continuum of H II 
galaxies. 
Accordingly, we suggest that the $C_{3727} - C_{4861}$ colour index 
provides a valid criterium to separate AGNs from H II galaxies for local 
galaxies. However, its behaviour at higher redshift is difficult 
to predict, specially for the case of AGNs. Therefore,
we advise the reader to be cautious when applying this index to 
redshifted ELGs. 

It is possible that similar results may be found by using other continuum 
colour indices (e.g.~Kennicutt 1992). We have used the [O II]$\lambda$3727 $\AA$\
and H$\beta$ lines
because these were the ones for which we maximized the number of ELGs with 
observed equivalent widths.


\begin{figure}
{\psfig{file=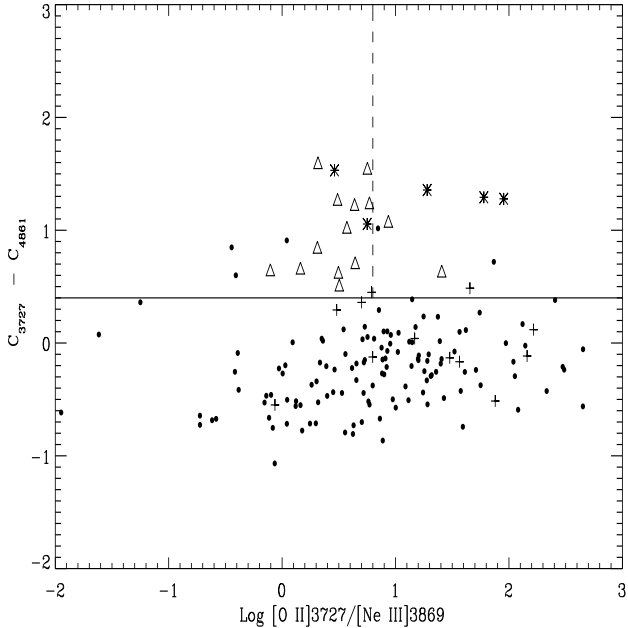,width=9.0cm,height=9.0cm}}
\vspace{-0.3cm}
\caption{\label{ffa}   
[O II]$\lambda$3727/[Ne III]$\lambda$3869 intensity ratio versus 
$C_{3727} - C_{4861}$. Symbols as in Figure~1. The data are not reddening 
corrected. Reasonably good segregation is attained between
LINERs and Seyferts 2 (for 
$\log$~([O II]$\lambda$3727/[Ne III]$\lambda$3869) $>$ 0.8 and 
$C_{3727} - C_{4861}$ $>$ 0.4, approximately).}
\end{figure}

\subsection {Emission-line ratio diagrams}

The strongest emission-lines observed in the near ultraviolet part of the 
spectrum of ELGs
are [O II]$\lambda$3727 $\AA$\ and [Ne III]$\lambda$3869 $\AA$.

The oxygen doublet [O II]$\lambda$3727 $\AA$\ is particularly intense in objects which
contain a large, partially ionised, transition region, like is the case 
in active galaxies. In this region, high energy photoelectrons, resulting 
from photoionisation by X-ray photons and by 
Auger processes, collisionally excite the O$^{+}$ ion, along with
other atoms/ions like O$^{0}$, S$^{+}$ and N$^{+}$.

The  neon [Ne III]$\lambda$3869 $\AA$\ line is usually more conspicuous in AGNs 
than in H II galaxies. This may be due to the lack of strong metallic 
absorption edges in the AGNs ionising continuum compared with the atmospheres 
of the stars exciting the H II regions where the formation of Ne$^{+2}$ 
tends to be suppressed relative to O$^{+2}$ (Balick \& Snedon, 1976; 
Shields \& Searle, 1978). Additionally, the fact that the ionisation 
potentials of O$^{+2}$ and Ne$^{+2}$ are relatively close (35.1 eV and 41 eV, 
respectively) and that [Ne III]$\lambda$3869 $\AA$\ is also a nebular line, 
suggests  that [O III]$\lambda$5007 $\AA$\
can be replaced by [Ne III]$\lambda$3869 $\AA$\
in the diagnostic diagrams of figure~\ref{f0}.

Replacing [O III]$\lambda$5007 $\AA$\ by the [Ne III]$\lambda$3869 $\AA$\ line in the 
[O II]$\lambda$3727/H$\beta$ against [O III]$\lambda$5007/H$\beta$ diagnostic
diagram of figure~\ref{f0}, we obtained a rather similar diagram (although
the new diagram is not corrected for reddening and it depends on
the Ne$/$O abundance ratio), which is displayed in figure~\ref{fa} 
(left diagram).
This new diagram seems also quite efficient in separating active galaxies
from H II galaxies. In fact, there is an upper limit situated at 
$\log$ ([Ne III]$\lambda$3869/H$\beta$)\footnote{intensity ratio not reddening 
corrected.} $\approx$ - 0.2, above which no H II galaxies are observed. 

LINERs are quite difficult to separate from Seyferts 2 with so few adequate 
diagnostic lines as available from deep redshift surveys. Nevertheless, the
[O III]$\lambda$5007/H$\beta$ intensity ratio (see the right panel of 
figure~\ref{fa}) allows some level of segregation  between LINERs 
($\log$~([O III]$\lambda$5007$/$H$\beta$) $\leq $ 0.5) and Seyferts 2 
($\log$~([O III]$\lambda$5007$/$H$\beta$) $> $ 0.5).
Furthermore, if we now plot the $C_{3727} - C_{4861}$ colour index against the
[O II]$\lambda$3727$/$[Ne III]$\lambda$3869 line ratio, without correcting 
them for reddening (see figure~\ref{ffa}),
we can define three regions in the diagram: one for H II galaxies (the bottom
region), one for LINERs (the upper right one) and one for LINERs and Seyferts 2
(the upper left one).  We have to be cautious in our conclusions because
of the very small sample of LINERs available to us, but it seems 
that the region limited by 
$\log$~([O II]$\lambda$3727$/$[Ne III]$\lambda$3869) $>$ 0.8 and 
$C_{3727} - C_{4861}$ $>$ 0.4, approximately, tends to be more 
populated by LINERs than by the other two types of ELGs.

\subsection{Equivalent width diagrams}

All the diagrams considered up to now involve the use of emission-line or 
continuum intensities. 
Unfortunately, in many of the existent redshift survey data bases the 
observed spectra are not flux calibrated, which difficults the use
of these diagrams. For this reason, we decided to determine other
diagnostic diagrams which would overcome this problem by
using only equivalent widths. 
\begin{figure*} 
\centerline{\psfig{file=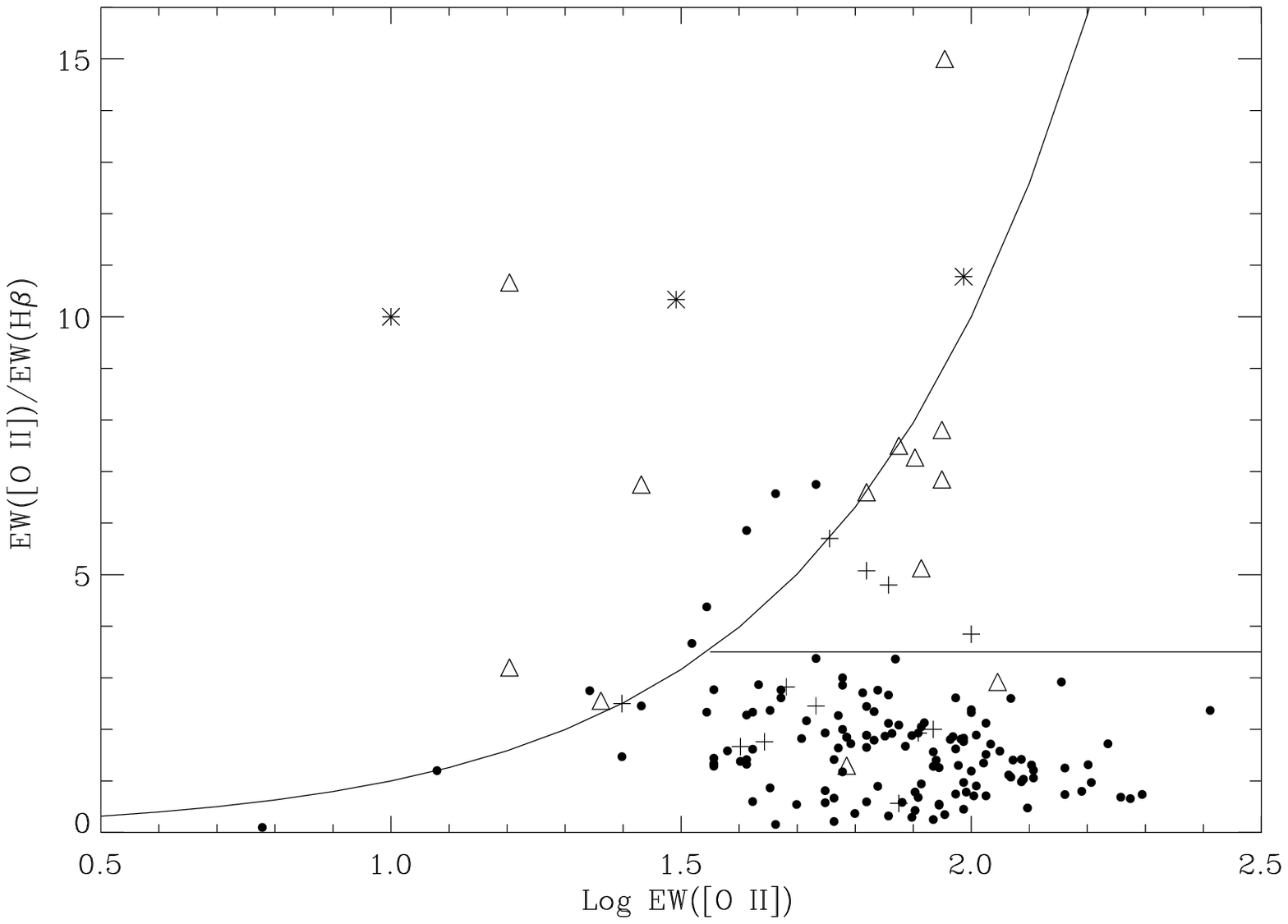,width=9.0cm,height=9.0cm}
\hspace{-0.3cm}\psfig{file=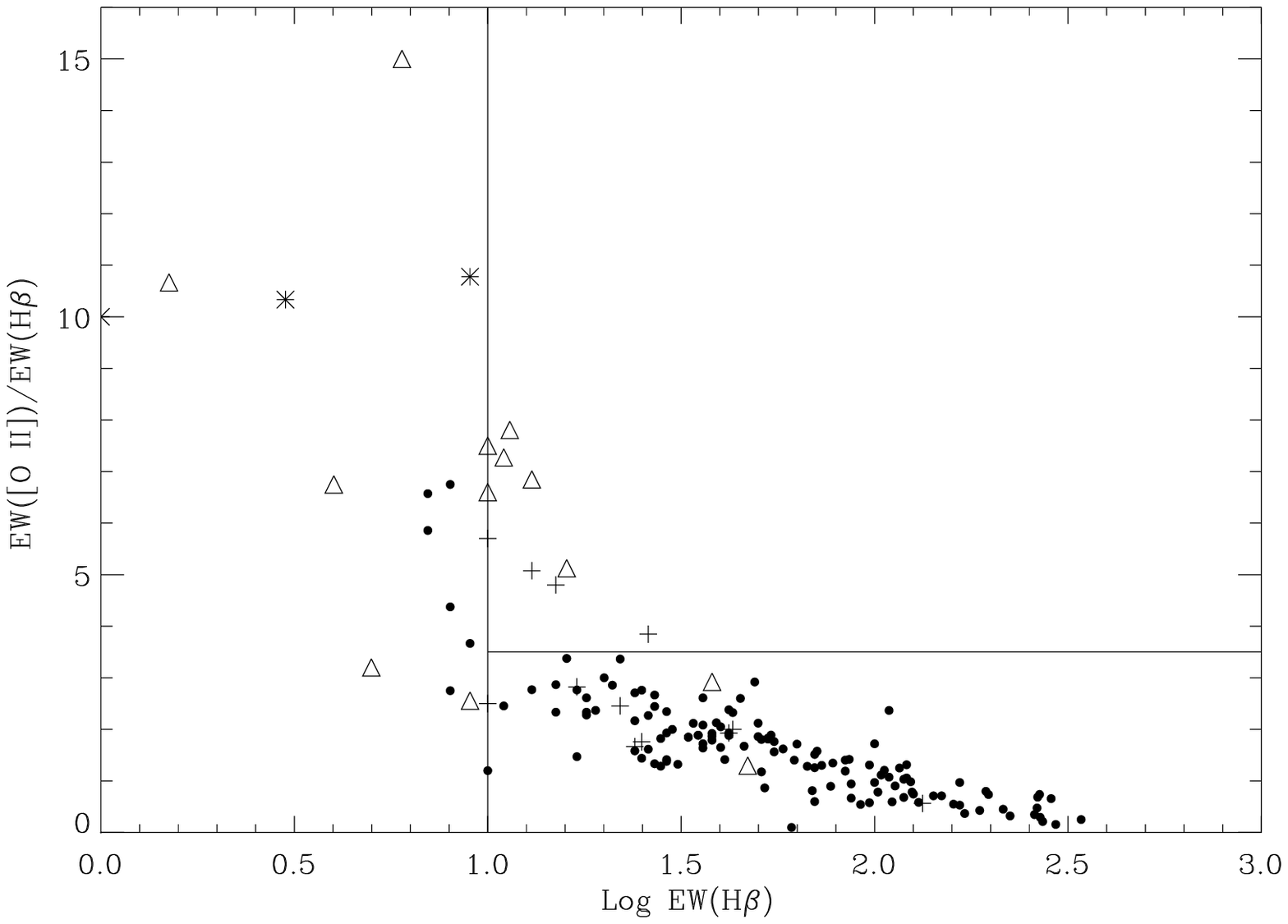,width=9.0cm,height=9.0cm}}
\vspace{-0.3cm}
\caption{\label{f5} EW([O II]$\lambda$3727)$/$EW(H$\beta$) ratio
versus EW([O II]$\lambda$3727 (left diagram) and versus EW(H$\beta$) (right 
diagram).  In the left diagram, the curve represents
the points with EW(H$\beta$) equal to 10 $\AA$, and objects 
above it have a lower value. The horizontal line was drawn based on 
the distribution of the different types of galaxies as a function of 
the EW([O II]$\lambda$3727)$/$EW(H$\beta$) ratio, so that most of our
active galaxies are located above 3.5. The same separation limits 
apply to the diagram on the right. Symbols are the same as in Figure~1.}
\end{figure*}

In  figure~\ref{f5} we plotted  EW([O II]$\lambda$3727)$/$EW(H$\beta$) 
against EW([O II]) and  against EW(H$\beta$), on the left and right 
diagrams, respectively. The lines drawn in each diagram define 
three zones identifying the nature of the ELGs. The horizontal 
lines were defined according to the distribution  of 
the EW([O II]$\lambda$3727)$/$EW(H$\beta$) ratio. It is found 
that about 73 $\%$ of all the active galaxies in the  sample 
(against 8$\%$ of all H II galaxies) are 
located above  the value of 3.5. 
The curve and vertical lines in the left and right diagrams, respectively, 
are based on our data analysis where we have found that an 
EW(H$\beta$) of 10 $\AA$ is a good upper limit to
separate active galaxies from H II galaxies as only a 
few H II galaxies lie below this limit. 
We have tested the validity of this limit, specially to check
if our H II galaxies sample is not biased towards
the blue, luminous, young H II galaxies ignoring the 
ones with low EW(H$\beta$) values. We have concluded that this is not 
the case, the blue, luminous H II galaxies are 
randomly distributed in terms of their EW(H$\beta$) values.

Thus, the combination of the two above mentioned limits leads to 
the following separation zones in both diagrams of 
Figure~\ref{f5}. The first one corresponds to the lower right region 
(henceforth called {\it H II galaxies region}), where 
most of the H II galaxies and  only a few  active ones
are located. It is limited upwards by 
the curve at EW([O II]$\lambda$3727)$/$EW(H$\beta$) $=$ 3.5 and
leftwards by the curve EW(H$\beta$) $=$ 10 $\AA$.
 About 87 $\%$ of all H II galaxies in our sample are 
located here against 15 $\%$ of all Seyferts 2 and no 
LINERs (which represent about 12 $\%$ of all active galaxies in the data 
sample).

The second one (hereafter called {\it AGN region}) corresponds 
to the sum of two zones. One is defined by the region where EW(H$\beta$) 
$\leq$ 10 $\AA$ (left region in diagrams).  The other corresponds to the 
location where EW([O II]$\lambda$3727)$/$EW(H$\beta$)
$\geq$ 3.5 {\it and}  EW(H$\beta$) 
$>$ 10 $\AA$ (upper right region in diagrams). 
About 85 $\%$ of all our Seyferts 2 and 100 $\%$ of all 
our LINERs are located here (about 88 $\%$ of all our 
active galaxies)  against 13 $\%$ of all H II galaxies.

Let us call these percentages  ``normalised frequencies'', as they  
are calculated as the frequency of ELGs of a certain type, in a certain region,  normalised to the 
total
number of ELGs of that type.

Considering that for each {\it region} the sum of 
the normalised frequencies, i. e.,  the normalised frequency of 
AGNs (${\rm Freq}$$_{\rm region}$(AGN)) 
{\it plus} the normalised frequency of 
H II galaxies (${\rm Freq}$$_{\rm region}$(H II)), is equal to unity,
we have made a crude estimate of the ``probability'' that an ELG which falls
in a given {\it region} of the diagnostic diagram is either an AGN or an 
H II galaxy. This illustrates the goodness
of the diagrams for separating the various types of ELGs. 
Although from our literature search we believe that 
our sample is representative of each type of ELG in the local Universe we 
should keep in mind that these ``probabilities'' could slightly change if, 
for example, we add more AGNs to our reference sample. Nevertheless and for the
sake of simplicity we will still call these  
{\it probabilities}.

This leads to determine that in the AGN region, the 
probability that an ELG is an AGN is
${\rm Prob}$$_{\rm AGN}$(AGN) $\approx$ 87 $\%$, and
that of it being an H II galaxy is ${\rm Prob}$$_{\rm AGN}$(H II) $\approx$ 
13 $\%$, 
while in the H II galaxies region,
${\rm Prob}$$_{\rm H \;II}$(AGN) $\approx$ 12 $\%$ and
${\rm Prob}$$_{\rm H \;II}$(H II) $\approx$ 88 $\%$.

Most LINERs  
appear in the left region of the diagrams of 
Figure~\ref{f5}\footnote{Note that some LINERs are located up,  
out of the visible plotted range in the diagrams.}
 i. e., where  EW(H$\beta$) is smaller than 10 and that we call
{\it region AGN I}.
Can we use  these diagrams to separate LINERs from 
Seyfert 2 galaxies?
In {\it region AGN I}  we find about 11 $\%$ of all H II 
galaxies and about 53 $\%$ of all active galaxies,
which corresponds to 86 $\%$  of all the LINERs and 44 
$\%$ of all Seyferts 2. In terms of probabilities we have that 
in {\it region AGN I}: 
${\rm Prob}$$_{\rm AGN \;I}$(Seyf. 2) $\approx$ 32 $\%$,
${\rm Prob}$$_{\rm AGN \;I}$(LINER) $\approx$ 61 $\%$ and 
${\rm Prob}$$_{\rm AGN \;I}$(H II) $\approx$ 7 $\%$.
However, we are dealing with a
small number of LINERs which may not be representative 
of the whole LINERs family.

{\it Region AGN II} is the top/right one in the diagrams of Figure~\ref{f5}, limited 
to those ELGs where
EW([O II]$\lambda$3727)$/$EW(H$\beta$) $\geq$ 3.5 and 
EW(H$\beta$) $>$ 10.  About 
35 $\%$ of all active galaxies (corresponding to 41 $\%$ of all the 
Seyfert 2 galaxies and 14 $\%$ of all LINERs) and 1 $\%$ of all H II 
galaxies are located here.
In terms of probabilities: 
${\rm Prob}$$_{\rm AGN \;II}$(Seyf. 2) $\approx$ 75 $\%$,
${\rm Prob}$$_{\rm AGN \;II}$(LINER) $\approx$ 25 $\%$ and 
${\rm Prob}$$_{\rm AGN \;II}$(H II) $=$ 0 $\%$. 
Therefore, this is the region where the probability 
that an ELG is a Seyfert 2 is maximum.

\subsection{Transition/Ambiguous emission-line galaxies}

It is important to also consider the behaviour of the 12 transition/ambiguous 
ELGs of the reference sample (recall subsection~3.3) in the new diagrams. 
In all of them, from figure~3 to figure~5, 
these ELGs are located mainly in the H II galaxies region. 
Although in the equivalent width diagrams of Figure~5 few  of the 
transition/ambiguous objects are 
placed in the AGN II region, they are all close to the separation limits.

The $C_{3727} - C_{4861}$  colour index distribution of
the 12 ambiguous ELGs, range from approximately $-0.6$ to $0.4$ with the  mean 
value 0.01, shows a large superposition  with that of the H II galaxies.

Thus, the transition/ambiguous ELGs are located  mainly with the 
H II galaxies in the new diagrams with little spillage over the AGN region. 
This could lead essentially to three possibilities: 1) these objects are in fact H II/AGN composite, with an AGN component  weaker than the emission component produced from 
ionisation by stellar sources; 2) they are H II galaxies with a peculiar hot stellar population; 3) these galaxies have just different properties in their ionized regions. It is difficult to determine what makes these objects different from H II galaxies; future work will have to address these points.

Nevertheless, a clear conclusion can be drawn: the distribution of the
unclassified ELGs strengthens the validity of our new methods
for redshift surveys, in the sense that the possibility of finding a 
transition/ambiguous object in the AGN region of the diagrams is almost
negligible.

\subsection{Evolutionary effects and application to CFRS sample}

Galaxy evolution with redshift may
possibly modify the stellar population and gas structure
and composition of ELGs, affecting somehow the $C_{3727} - C_{4861}$  colour 
index, but will not strongly affect  emission-line intensity ratios as they 
depend only on the nebular emission.

It is difficult to predict both how the continuum
will evolve and  how evolution will affect the equivalent widths of the 
[O II]$\lambda$3727 $\AA$\ and H$\beta$ lines. In fact, theoretical 
models trying to 
represent evolving H II galaxies do not generally consider important 
details as the stellar absorption under H$\beta$, 
while for Seyferts 2 and LINERs the mechanisms responsible for the gas 
ionisation are not well known. Nevertheless, it is important to determine the 
effect of evolution in the continuum and in the equivalent width of these 
emission-lines in order to understand the physics 
of the ELGs observed in intermediate and high redshift surveys.
This will have to be studied in detail in future work. 

In the mean time, we can take an empirical  approach and compare the values 
of the  equivalent widths of the 
[O II]$\lambda$3727 $\AA$\ and H$\beta$ lines in our sample with the ones obtained 
in deep redshift surveys (e.g., LDSS: Colless 
{\it et al.} 1990 or CFRS: Tresse {\it et al.} 1996). We notice (see figure~6 
below) an absence of survey objects with EW(H$\beta$) larger than 
$\approx $ 55 $\AA$\ while in the local sample of H II galaxies, the  
EW(H$\beta$) reaches values up to $\approx $ 320 $\AA$.
This could be a real effect, evidence for galaxy evolution with redshift, or
an artifact due to observational selection effects attributed to either
instrumentation or most probably to the way samples were selected. At 
this stage it is impossible to answer these questions.
Nevertheless, this is an important issue which deserves future investigation.

\begin{figure}
\vspace{-4.5cm}
{\hspace{-1.1cm}{\psfig{file=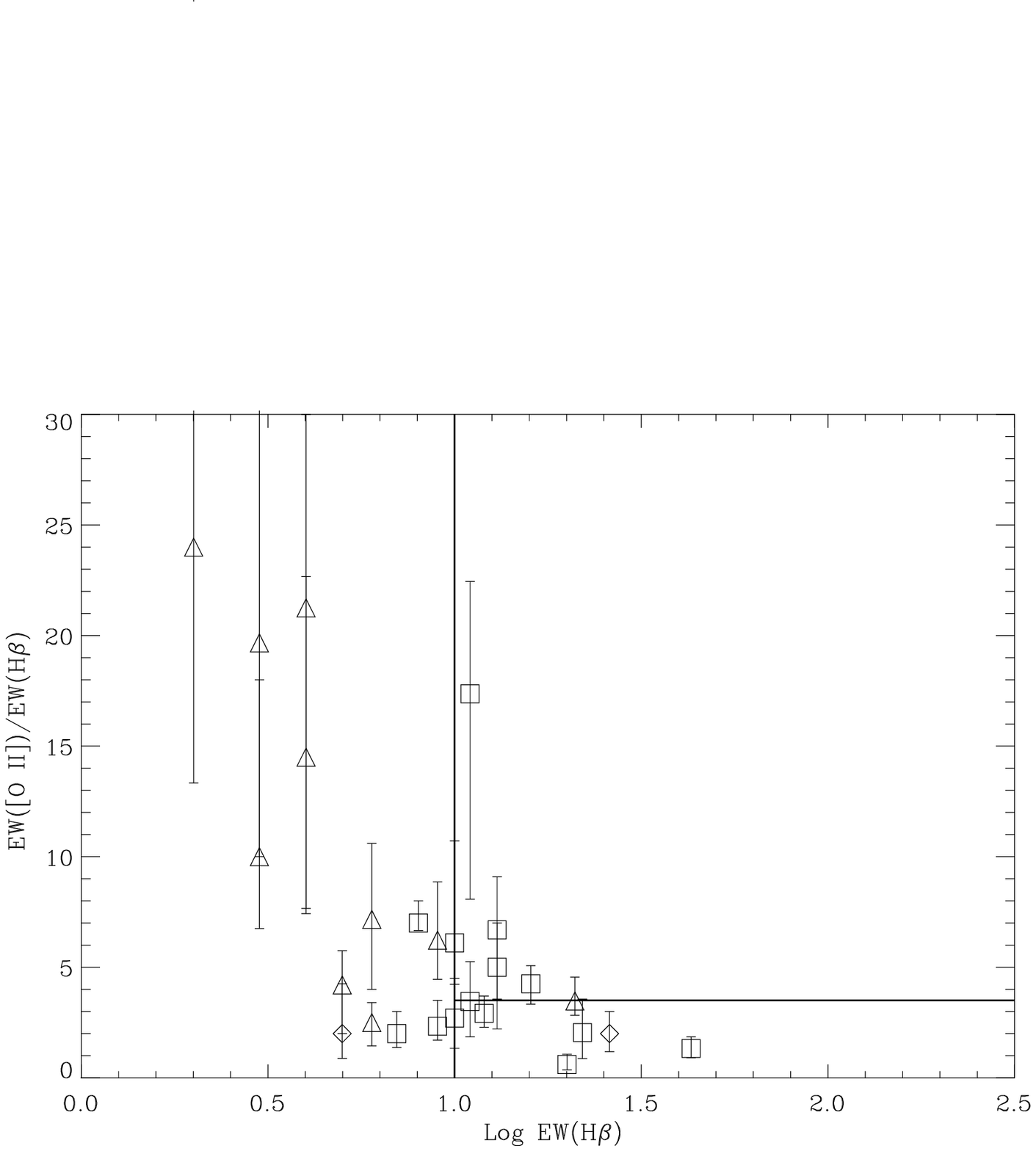,width=9.5cm,height=14.0cm,angle=-90}}}
\vspace{-0.3cm}
\caption{\label{CFRS} CFRS data plotted in our diagnostic diagram 
EW([O II]$\lambda$3727)$/$EW(H$\beta$) versus $\log$~EW(H$\beta$).
The classification is that given by
Tresse {\it et al.} (1996): squares are H II galaxies, triangles are 
active galaxies and diamonds are intermediate (or transition) objects.}
\end{figure}

In order to test whether the results obtained with the new diagrams of 
figure~5 
are consistent with the ones determined from the standard diagnostic methods,
we used the former to classify the CFRS ELGs analysed by Tresse {\it et al.} 
1996 (i.e.~those with $z \leq 0.3$).
We have plotted in figure~\ref{CFRS} the corresponding sub-sample of CFRS ELGs,
for which [O II]$\lambda$3727 $\AA$\ and H$\beta$ have been observed, and 
their equivalent widths have been measured. 
Tresse {\it et al.} (1996) have used standard diagnostic diagrams to separate 
the AGN from the H II ELGs in their sample. Using our figure~5  diagrams 
we have found that about 33 $\%$ of the galaxies up to $z = 0.3$ are active galaxies.
Due to the large abscissa error bars,\footnote{For the sake of clarity, we
did not plot the abscissa error bars in figure~6.} 7 objects are compatible 
both with being an AGN and an H II galaxy. Excluding them, the fraction of AGNs falls down to
about 20 $\%$, which is in quite good agreement with the 
(mean) 17 $\%$ fraction 
obtained by Tresse {\it et al.} (1996) using the classical diagnostic 
diagrams based only on emission-line intensity ratios. 
Furthermore, if we compare our classification for each CFRS z $\leq$ 0.3 ELG with the one of 
Tresse {\it et al.} (1996),  we check that these disagree only for 4 objects.

Unfortunately, due to the large error bars in 
the Tresse {\it et al.} (1996) sample, it is  not possible to draw
any conclusions regarding the $C_{3727} - C_{4861}$ colour index.

\section{Discussion and Conclusion}

The main goal of this work was to investigate new diagnostic methods
appropriate to classify the ELGs observed in deep redshift surveys. 
Most of the more  prominent optical lines used in the standard diagnostic 
diagrams (Baldwin, Philips \& Terlevich 1985; Veilleux \& Osterbrock 1987) 
move out of the observable optical spectral range for objects with $z > 0.3$.
In fact, for $z > 0.3$ the only strong emission-lines remaining in 
the optical range are H$\beta$ and [O III]$\lambda\lambda$4959, 5007 $\AA$,
observable up to $z \approx $0.7, and [O II]$\lambda$3727 $\AA$\ and  
[Ne III]$\lambda$3869 $\AA$\ observable up to $z \approx $1.2. 
This implies that the methods used to classify the  ELGs observed in redshift 
surveys will have to be independent of reddening  and to rely on a minimum of 
information from the optical/near ultraviolet part of the spectrum. 
Hence, we based our study mainly on the [O II]$\lambda$3727 $\AA$, 
[Ne III]$\lambda$3869 $\AA$, H$\beta$ and [O III]$\lambda$5007 $\AA$\
emission-lines and on 
a new continuum colour index. All of these can be observed with CCD detectors  
up to a redshift of $\sim 0.75$. 

Using as a reference set a large number of nearby ELGs, 
we have found that:

a)  The $C_{3727} - C_{4861}$ colour index
provides a good discriminator between local H II galaxies and active galaxies, 
i.e. Seyferts 2 and LINERs, while the [O II]$\lambda$3727 $\AA$\ to
[Ne III]$\lambda$3869 $\AA$\ line intensity ratio segregates between Seyferts 2
and LINERs. 
Nevertheless, $C_{3727} - C_{4861}$ should be used with caution for 
higher redshift galaxies as its behaviour with redshift is difficult
to predict, specially for the case of active galaxies. 

b) Diagrams using the [Ne III]$\lambda$3869/H$\beta$, 
[O II]$\lambda$3727/H$\beta$ and [O III]$\lambda$5007/H$\beta$ intensity 
ratios provide also reliable classification for ELGs. These have the 
advantage of depending only on the nebular emission.

c) Good discrimination between AGNs and H II galaxies is provided by the 
EW([O II]$\lambda$3727)$/$EW(H$\beta$) versus $\log$~EW(H$\beta$)
and EW([O II]$\lambda$3727)$/$EW(H$\beta$) versus
$\log$~EW([O II]$\lambda$3727) diagnostic diagrams. The use of equivalent 
widths instead of line ratios is 
particularly useful when the ELGs spectra are not flux calibrated. 
This method was tested using the Canada-France Redshift Survey  
$0 < z \leq 0.3$ ELGs sample. A good agreement was found between the 
classification obtained with this method and the results determined using the 
standard diagnostic diagrams (Tresse {\it et al.} 1996).

\vspace{0.5cm}

Although the level of discrimination obtained by the new diagnostic diagrams 
seems good, we have to be careful when aplying them to high redshift 
samples.  In the case of H II galaxies it is
complicated to accurately predict how galaxy evolution will affect both the 
$C_{3727} - C_{4861}$
colour index and  the equivalent widths of the [O II]$\lambda$3727 $\AA$\
and H$\beta$ lines  beyond $z = 0.3$.  For AGNs it is just not possible
to predict their spectral evolution  
with look-back time, as 
the mechanisms responsible for the energy generation and  
ionisation in these objects are still not well understood.  

Undoubtedly, following work will bring a clearer view on the correlation of 
these
observable quantities with the physical phenomena occurring in distant ELGs.  
Furthermore, it will be very important to be able to predict
the evolution of all the observable spectral quantities with
redshift,  for H II galaxies, Seyferts 2 and LINERs, taking into account both 
stellar and nebular contributions to the observed integrated spectra.
The application of these new methods to the large database of optical/near 
UV spectra of ELGs with $0 < $ $z \leq$ 0.7 -- 0.8 currently
available will allow to take a step forward in this
direction and into the physical interpretation of galaxy evolution 
with look-back time.   

\section*{Acknowledgments}

Discussions with Timothy Heckman, Richard Ellis, Robert Kennicutt, Joseph 
Shields, Fran\c cois Hammer, Guillermo Tenorio-Tagle, Luis Ho and Il\' \i dio Lopes are gratefully 
acknowledged. CR thanks the support of the ``Junta Nacional de 
Investiga\c c\~ao Cient\' \i fica e Tecnol\'ogica'' (Portugal) through a
grant BPD/6064/95, PRAXIS XXI program.
ET thanks the hospitality of the Royal Greenwich Observatory.
Financial support from an ANTARES (E.C.) grant is also acknowledged.

\appendix 
 
\bsp 
 
\end{document}